\documentstyle[stwol,epsfig]{article}

\def\be{\begin{equation}}
\def\ee{\end{equation}}
\def\bea{\begin{eqnarray}}
\def\eea{\end{eqnarray}}
\begin{document}

\title{A simple approach to describe hadron production rates
       in $\mbox{e}^{+}\mbox{e}^{-}$ annihilation~$^{*}$}

\author{ Yi-Jin Pei }

\address{$^{1}$CERN, CH-1211 Geneva 23, Switzerland (e-mail: peiyj@vxcern.cern.ch)\\
         $^{2}$I.~Physikalisches~Institut der RWTH, D-52074 Aachen, Germany~~~~~~~~~~}


\twocolumn[\maketitle\abstracts{
We show that, based on the idea of string fragmentation, the production 
rates of light flavored mesons and baryons originating from fragmentation can be
described by the spin,  the binding energy of the particle, and a 
strangeness suppression factor. 
Apart from a normalization factor, $\mbox{e}^{+}\mbox{e}^{-}$ 
data at different center-of-mass energies can be described simultaneously. 
Applying to the heavy flavor production, we find that our predictions
are in good agreement with data.}]

\section{Introduction}
\vspace*{-0.1cm}
The soft processes of hadronization are not calculable with a perturbative 
approach and instead rely currently on phenomenological models.
The most successful of these
models are the string~\cite{string} and the cluster~\cite{cluster}
fragmentation models, implemented in the Monte Carlo programs
JETSET~\cite{JETSET} and HERWIG~\cite{HERWIG}, respectively. 
However, these models either require a large number of free parameters
in order to reproduce the measured hadron production rates 
(JETSET), or do not give a satisfactory 
description of baryon data (HERWIG).
A review may be found in Ref.~\cite{sj1,lep2rev}.

\footnotetext{$^{*}$~Presented at ICHEP'96, July, 1996, Warsaw, Poland}

In this paper, we first show some regularities in the hadron production rates 
measured at LEP. Then based on the idea of string fragmentation, we 
deduce a simple formula to describe the LEP data. The formula is also
applied to data obtained at center-of-mass energies around 10~GeV and
29-35~GeV, and to the heavy flavor production. 

\vspace*{-0.1cm}

\section{Overview of LEP data}
\vspace*{-0.1cm}
Thanks to excellent performance of the detectors and high statistics available, 
very careful work by all four LEP experiments has given a very complete
picture of  the production of identified particles from 
$\mbox{e}^{+}\mbox{e}^{-}$ annihilation. 
The measured production rates  per hadronic Z 
event for the identified particles at LEP~\cite{QGYPange,Lambda}
are listed in Table~\ref{QGYPrate}. 
%For all mesons  and octet baryons 
%the measurements are in good agreement between experiments.
%For the decuplet baryons, however, there are still  discrepancies 
%between experiments, reflecting difficulties in the measurements. 

Studies of general features of particle production, such as the fraction 
$V/(V+P)$ of mesons produced in spin-1 states and the strangeness 
suppression factor $\gamma_{s}=\mbox{s}/\mbox{u}$, provide useful information about 
the fragmentation process. The ratio $V/(V+P)$ for primary mesons is expected 
to be equal to 0.75 from simple spin counting. 
%We define primary hadrons
%as those which are not decay products of other hadrons, i.e. as hadrons
%originating from fragmentation or containing a primary quark $q$ from the
%process $\mbox{e}^{+}\mbox{e}^{-}\rightarrow q\bar{q}$.
From the rates and the primary fractions given in Table~\ref{QGYPrate} and 
Ref.~\cite{peippe}, 
%after subtracting the decay contributions for the light flavored mesons by 
%using the fraction values listed in the last column of Table~\ref{QGYPrate},
%we obtain:
%
%{\footnotesize
%\begin{eqnarray*} 
%\mbox{Mesons~~~~~~~~~~~~~~~~~~~~~~~~}      & \mbox{~~~~~}  V/(V+P) \\
%\rho^{0}, \omega, \pi^{0}, \eta , \eta^{\prime} \mbox{~} (\Delta m_{\rho - \pi} =0.63\mbox{~GeV}) 
%                           & \mbox{~~~~~} 0.46\pm 0.04 \\
%\mbox{K}^{*}, \mbox{K} \mbox{~} (\Delta m_{\mbox{{\tiny K}}^{*} - \mbox{{\tiny K}}} =0.40\mbox{~GeV})
%  \mbox{~~~~~~~}                          & \mbox{~~~~~} 0.42 \pm 0.03 \\   
%\mbox{D}^{*}, \mbox{D} \mbox{~}(\Delta m_{\mbox{{\tiny D}}^{*} - \mbox{{\tiny D}}} =0.14\mbox{~GeV})
% \mbox{~~~~~~~}                           & \mbox{~~~~~} 0.56 \pm 0.04 \\
%\mbox{B}^{*}, \mbox{B} \mbox{~}(\Delta m_{\mbox{{\tiny B}}^{*} - \mbox{{\tiny B}}} =0.046\mbox{~GeV})  
% \mbox{~~~~~~}                           & \mbox{~~~~~} 0.75 \pm 0.04    
%\end{eqnarray*} 
%} 
we obtain a value of $0.46\pm 0.04$, $0.42 \pm 0.03$, $0.56 \pm 0.04$ and
$0.75 \pm 0.04$ for u(d)-, s-, c- and b-mesons respectively.
The low value of $V/(V+P)$ for light flavored mesons could be explained 
by mass differences between the vector
and pseudoscalar mesons, i.e. by the relatively larger binding energy of
pseudoscalar mesons. 
For c-mesons the measured ratio of $V/(V+P)$  is also low, while 
for b-mesons it agrees well with  the expected value of 0.75. 
We will discuss this later in more detail.

Figure~1 shows the production rates of primary had\-rons
(measured rate~$\times$~fraction from JETSET as listed in Table~\ref{QGYPrate}),
divided by the spin factor (2J+1), as a function of the hadron mass.
We see similar behavior~\footnote{From Fig.~1 one can see that it
is not possible to describe all the data with just an exponential  function of
the hadron mass (or mass squared).} (steps)
for hadrons belonging to the same
multiplet, such as $(\rho /\omega,\mbox{K}^{*},\phi)$, $(\mbox{p},\Lambda,\Sigma,
\Xi )$, $(\Delta , \Sigma^{*}, \Xi^{*} ,\Omega )$ and 
$(\mbox{f}_{2}$,$\mbox{K}_{2}^{*}$,$\mbox{f}_{2}^{\prime})$. 
The mass difference of  hadron pairs which are in the same
multiplet but differ by one in the strangeness 
is in the range of 0.1--0.15~GeV  (except $m_{\mbox{{\tiny K}}}$$-$$m_{\pi}$).
This is close to  the mass difference between the s and the u(d) quark,
showing that the binding energy of hadrons in each pair is about the same.
Unlike the case of $V/(V+P)$, one could expect the $\gamma_{s}$
value determined from different hadron pairs is about the same.

In the string fragmentation model, one  expects 
the strangeness suppression factor $\gamma_{s}$ to be around 0.3. 
This parameter can be measured from the production rates of strange compared with
non-strange hadrons, and from the momentum spectrum of strange mesons.
The results are very consistent with the expectation 
(a review may be found in Ref.~\cite{peippe}). 
%Also,  data from p$\bar{\mbox{p}}$ 
%and pp collisions (see e.g. Ref.~\cite{ua2}) give a  
%$\gamma_{s}$ value around the expectation~\footnote{However, 
%neutrino experiments at lower energies~\cite{QGYPs2u1}, and recently, 
%both ZEUS~\cite{QGYPs2u2} and H1~\cite{QGYPs2u3}, 
%require a lower value of about 0.2 for $\gamma_{s}$. More careful 
%studies in this area are needed in the future.}.
This suggests that the strangeness suppression occurs at the quark level. 
In the following we consider that the particle production proceeds in two
stages, namely quark pair production in the color string field and 
successive recombination.

\begin{table}[t]
\vspace*{-0.4cm} 
{\scriptsize
%\begin{center}
\caption{Average particle production rates
         (excluding charge conjugates and antiparticles), 
         compared to the calculated values (see next section). 
         The fraction of primary hadrons 
         obtained from the fit and JETSET is also shown.}
\vspace{0.2cm}
\begin{tabular}{|lllll|} \hline   
 Particle    &  Rate              &  Rate      & Prim.     & Prim.   \\   
             &  Measured          &  Calc.     & Frac.     & Frac.   \\    
             &                    &            & Calc.     & {\tiny JETSET}  \\ \hline   
$\pi^{0}$      & $9.19\pm 0.73$   &  9.77    &    0.16     &  0.14        \\ 
$\pi^{+}$      & $8.53\pm 0.22$   &  8.70    &    0.18     &  0.16        \\ 
$\mbox{K}^{0}$ & $1.006\pm 0.017$ &  1.008   &    0.25     &  0.30        \\ 
$\mbox{K}^{+}$ & $1.185\pm 0.065$ &  1.100   &    0.23     &  0.27        \\ 
$\eta$         & $0.95\pm 0.11$   &  0.85    &    0.30     &  0.33         \\ 
$\eta^{\prime}$& $0.22\pm 0.07$   &  0.11    &    0.59     &  0.79        \\ \hline 
$\mbox{f}_{0}(980)$ & $0.140\pm 0.034$ & 0.080 &  0.99     &  0.93        \\ \hline 
$\rho^{0}$     & $1.29\pm 0.13$   & 1.12     &    0.47     &   0.54        \\ 
$\mbox{K}^{*0}$& $0.380\pm 0.021$ & 0.390    &    0.49     &   0.60        \\ 
$\mbox{K}^{*+}$& $0.358\pm 0.034$ & 0.395    &    0.50     &   0.60        \\ 
$\omega$       & $1.11\pm 0.14$   & 1.05     &    0.48     &   0.57        \\ 
$\phi$         & $0.107\pm 0.009$ & 0.092    &    0.64     &   0.70        \\ \hline 
$\mbox{f}_{2}(1270)$  & $0.25 \pm 0.08$  &  0.19  &  0.78  &   0.96      \\ 
$\mbox{K}^{*}_{2}(1430)^{0}$& $0.095\pm 0.035$ & 0.051    &  1.00  &  0.98    \\  
$\mbox{f}^{\prime}_{2}(1525)$ & $0.020\pm 0.008$ & 0.018 & 1.00  &  0.98    \\ \hline 
p              & $0.49\pm 0.05$     &  0.54     &   0.12    &     0.56      \\ 
$\Lambda$      & $0.186\pm 0.008$   &  0.165    &   0.12    &     0.44      \\ 
$\Sigma^{0}$   & $0.0355\pm 0.0065$ &  0.0387   &   0.40    &     0.86     \\ 
$\Sigma^{+}$   & $0.044\pm 0.006$   &  0.037    &   0.42    &     0.88     \\ 
$\Xi^{-}$      & $0.0129\pm 0.0007$ &  0.0118   &   0.41    &     0.75     \\ \hline 
$\Delta^{++}$  & $0.064\pm 0.033$     & 0.069   &   0.69    &     0.95      \\ 
$\Sigma(1385)^{+}$ & $0.011\pm 0.002$ & 0.016   &   0.91    &     0.92      \\ 
$\Xi(1530)^{0}$ & $0.0031\pm 0.0006$  & 0.0050  &   0.94    &     0.94      \\ 
$\Omega^{-}$   & $0.0008\pm 0.0003$   & 0.0015  &   0.88    &     0.92      \\ \hline 
$\Lambda(1520)$ & $0.0107\pm 0.0014$  & 0.0129  &   0.71    &     ---       \\ \hline 
\end{tabular}
\label{QGYPrate}
%\end{center}
}
\vspace*{-0.3cm} 
\end{table}

\begin{figure}
\vspace*{-0.9cm} 
\center
%\rule{2cm}{0.2mm}\hfill \rule{2cm}{0.2mm}
%\vskip 2.8cm
%\rule{2cm}{0.2mm}\hfill \rule{2cm}{0.2mm}
%\psfig{figure=/user/peiyj/tex/lep2/rate_mass_n.eps,height=1.5in}
%\psfig{figure=/user/peiyj/tex/lep2/rate_mass_n.eps,height=8.5cm}
\mbox{\epsfig{file=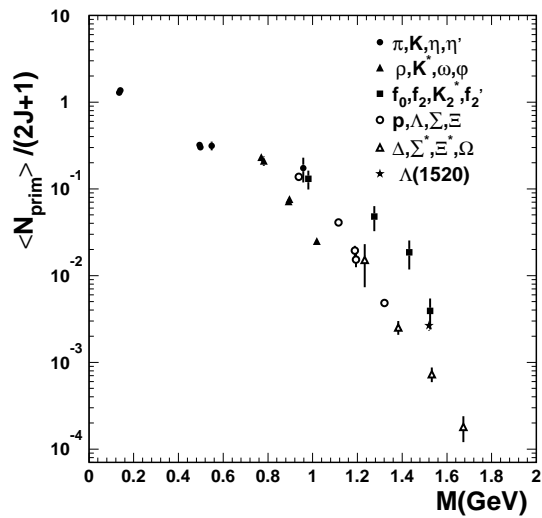,width=8.5cm}}
\vspace*{-1.7cm} 
\caption{Production rates of primary light flavored hadrons at LEP energy, 
divided by the spin factor (2J+1), as a function of the hadron mass.}
%\label{fig:radk}
\vspace*{-0.4cm} 
\end{figure}

\vspace*{-0.1cm}
\section{Analysis}
\vspace*{-0.1cm}
Quark pair production in the color string field
can be considered as a tunneling process. 
The probability to produce a $q\bar{q}$ pair 
is proportional to exp($-\pi m_{q}^{2}/\kappa$), 
where $m_{q}$ is the (constituent) quark mass, and $\kappa$ the string constant. 
We assume that the probability of quarks recombined to a hadron with the
mass $M_{h}$ is proportional to exp($-E_{bind}/T$), 
where $E_{bind}=M_{h}- \sum_{i} m_{q_{i}}$ is the hadron binding 
energy~\footnote{It can be ascribed to the hyperfine interaction~\cite{rosner}.},
and $T$ the effective temperature in hadronization. 
The production rates of light flavored
mesons and baryons from fragmentation can be described as 
\begin{equation}
 <N> = C \cdot \frac{2J+1}{C_{B}} \cdot (\gamma_{s})^{N_{s}}
       \cdot \mbox{e}^{-\frac{E_{bind}}{T}}, \label{eqrate}
\end{equation}
where $\gamma_{s}$$=$$\mbox{exp}(-\pi (m_{s}^{2}-m_{u}^{2})/\kappa )$ is
the strangeness suppression factor, $N_{s}$ the number of strange quarks
contained in the hadron, and $J$ the spin of the hadron. 
$C$ is an overall normalization factor, 
%which increases with the increasing center-of-mass 
%energy (reflecting the rise of multiplicities with increasing energy,
%which can be predicted by QCD~\cite{multi}).
and $C_{B}$ is the relative normalization
factor between mesons and baryons (for mesons $C_{B}$$=$$1$). 
%which should be independent on the center-of-mass energy.
Equation~\ref{eqrate} can also be applied to mixed states of the SU(3) octet
and singlet, such as $\eta $ and $\eta^{\prime}$, by adding up the
u$\bar{\mbox{u}}$(d$\bar{\mbox{d}}$) and s$\bar{\mbox{s}}$ contributions.
For this purpose we use the mixing formulae and angles given in Ref.~\cite{PDG}.

Total hadron rates  are calculated as follows. At first the 
number of light flavored hadrons produced from fragmentation is 
calculated by using  Eq.\ref{eqrate}. 
%Heavy quark production in the color string field
%is strongly suppressed due to the term exp($-\pi m_{q}^{2}/\kappa$)
%and can therefore be neglected.
For hadrons which contain a primary  quark $q$, 
we use Eq.\ref{eqrate} to determine their relative ratios, and then get
their rates by normalizing the sum of the rates  to the $q\bar{q}$ 
fraction, $\Gamma_{q\bar{q}}/\Gamma_{had}$, which can be
calculated  by the Standard Model.
All light flavored hadrons up to a mass of 2.5~GeV in the meson and
baryon summary table of Ref.~\cite{PDG} are included in the calculation. 
In the next step we let all these primary hadrons decay according to 
their decay channels and branching ratios given in Ref.~\cite{PDG}.
The decay chain stops when $\mu $, $\pi$, $\mbox{K}^{\pm}$, $\mbox{K}^{0}_{L}$
or stable particles are reached. 

In the fit we choose $\gamma_{s}$, $\Delta m=m_{s}-m_{u}$, $T$, $C$ and 
$C_{B}$ as free 
parameters~\footnote{Since exp$(-E_{bind}/T)=\mbox{exp}(\sum_{i} m_{u}/T)
\cdot \mbox{exp}(-(M_{h}-\sum_{i} (m_{q_{i}}-m_{u}) )/T)$,  the factor
$\mbox{exp}(\sum_{i} m_{u}/T)$ can be absorbed in $C$ and $C_{B}$.
The error function of the fit is mainly sensitive to the change in the 
mass difference  $m_{s}-m_{u}$.} 
(we assume $m_{u}$$=$$m_{d}$).
The results of the fit to the LEP data~\footnote{The new result for $\Lambda(1520)$
is not included in the fit.} 
for a typical s quark mass of 0.5~GeV are listed in 
Table~\ref{tabfit}. The calculated total rate 
and primary fraction for different hadrons are listed in Table~\ref{QGYPrate}.
We see a good agreement
between the measured and calculated rates except for some decuplet baryons.
%New results~\cite{Lambda,k2s} on $\Lambda(1520)$ and 
%$\mbox{K}^{*}_{2}(1430)^{0}$ agree also well with the calculated rates.
As mentioned in Ref.~\cite{QGYPange,lep2rev}, 
experimental errors for  the decuplet baryons
are still large and there are  discrepancies between experiments.
The large $\chi^{2}$ value of the fit is mainly due to the
decuplet baryons. If they are excluded in the fit,
the $\chi^{2}/dof$ is then equal to 22.6/15 while the fit results remain 
essentially unchanged.
However, the difference between data and our predictions for the 
decuplet baryons might suggest that baryon production is not described
properly in our approach,
since baryons are considerably more complicated objects than mesons.
%Our present description for baryons is consistent with the `popcorn'
%mechanism~\cite{popcorn}, according to which baryons appear from the successive
%production of several quark-antiquark pairs.
If diquark production~\cite{string} is the main mechanism for 
baryon production, it is then more suitable to use the effective diquark masses 
than the quark masses in Eq.\ref{eqrate}. This will, however, increase considerably
the number of  free parameters.

\begin{table}[t]
\vspace*{-0.4cm} 
\caption{Results of the fit to LEP data and to data obtained at 
                 different center-of-mass energies.}
\begin{center}
%\vspace{0.1cm}
{\scriptsize
\begin{tabular}{|l|l|l|} \hline   
 Parameters  & $\sqrt{s}=91$~GeV & Simultaneous Fit   \\ \hline\hline  
$\gamma_{s}$ & $0.29\pm 0.03$    & $0.29\pm 0.02$   \\ \hline   
$\Delta m$~(GeV)
             & $0.150\pm 0.029$  & $0.161\pm 0.024$ \\ \hline   
$T$~(GeV)    & $0.289\pm 0.020$  & $0.298\pm 0.015$ \\ \hline   
$C$          & $0.209\pm 0.041$  & $C_{91}=0.218\pm 0.034$ * \\ \hline   
$C_{B}$      &  $10.0\pm 1.0$    & $11.0\pm 0.9$    \\ \hline \hline   
$\chi^{2}/dof$ & 56.2/19         & 155.8/57         \\ \hline  
\multicolumn{3}{c} {*~ $C_{30}=0.124\pm 0.020$,
  ~~$C_{10} = 0.049\pm 0.008$}
\end{tabular}
}
\end{center}
\label{tabfit}
\vspace*{-0.4cm} 
\end{table}

%An important issue is whether our approach can also be applied 
%to data obtained at different  center-of-mass energies of 
%$\mbox{e}^{+}\mbox{e}^{-}$ annihilation. 
We  also perform a simultaneous fit to data at different
center-of-mass energies with 7 free parameters:
$\gamma_{s}$, $\Delta m$, $T$, $C_{B}$, $C_{91}$, $C_{30}$ and $C_{10}$,
where $C_{91}$, $C_{30}$ and $C_{10}$ are the overall normalization factor
at the center-of-mass energy 91, 29-35 and 10~GeV, respectively.
The fit results are listed in Table~\ref{tabfit} and shown in Fig.~2.
The value of $\gamma_{s}$, $\Delta m$, $T$ and $C_{B}$ is very consistent 
with that obtained from the LEP data alone, showing that the value of these
parameters is independent of the center-of-mass energy.

\begin{figure}
\vspace*{-0.9cm} 
\center
%\rule{2cm}{0.2mm}\hfill \rule{2cm}{0.2mm}
%\vskip 2.8cm
%\rule{2cm}{0.2mm}\hfill \rule{2cm}{0.2mm}
%\psfig{figure=/user/peiyj/tex/lep2/rate_mass_n.eps,height=1.5in}
%\psfig{figure=/user/peiyj/tex/lep2/rate_mass_n.eps,height=8.5cm}
\mbox{\epsfig{file=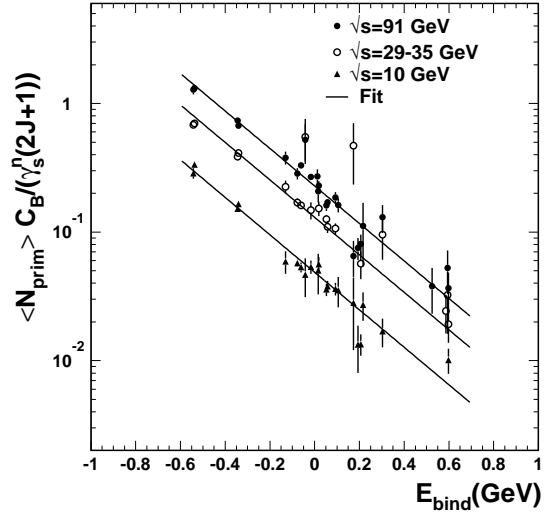,width=8.5cm}}
\vspace*{-1.6cm} 
\caption{Production rates of light flavored hadrons originating from fragmentation 
at different center-of-mass energies, multiplied by the factor 
$C_{B}/(\gamma_{s}^{N_{s}}(2J+1))$,
as a function of the binding energy of hadrons.}
%\label{fig:radk}
\vspace*{-0.4cm} 
\end{figure}

The $\gamma_{s}$ value determined from the fit is in an excellent agreement
with the expectation.
The value of $\Delta m=m_{s}-m_{u}$ required from the fit is  
consistent with that obtained from hadron masses~\cite{rosner}
and from QCD calculations~\cite{leut}.
While the values of $\gamma_{s}$, $\Delta m$ and $T$ do not depend
on the choice of the $m_{s}$ value,  
the values of $C$ and $C_{B}$ are correlated with the $m_{s}$ value. 
%They are changed~\footnote{If the current quark masses are used, 
%i.e. $m_{u,d}\simeq 5$MeV and $m_{s}\simeq 160$MeV, the value of $C_{B}$
%is around 3.}
%by about 20\% when $m_{u}$ and $m_{s}$  are shifted by 50~MeV. 
However, the calculated  baryon rates do not depend on the choice of 
the  $m_{s}$  value  since they are determined by both $m_{s}$ and $C_{B}$.
From our analysis we find that baryons contribute 10.9\%
of the primary produced hadrons at  LEP energy.

Fractions of primary hadrons determined from the fit are listed in 
Table~\ref{QGYPrate}. They are in general lower than those predicted
by JETSET due to the inclusion of orbitally excited states
up to a mass of 2.5~GeV in our analysis. JETSET includes some of 
orbitally excited meson states (only those with $L=1$),
but no orbitally excited baryon states. 
%(that is why the fractions for the octet baryons 
%determined from the fit are much lower than those from JETSET).
%As summarized in Ref.~\cite{lep2rev}, momentum spectra of light 
%flavored baryons predicted by JETSET (and other
%models) are too hard, which might be an indication for the existence of
%orbitally excited baryons. 
From our analysis we obtain the
fraction of primary produced states with orbital excitation
to be 27\% for mesons and 39\% for baryons  at  LEP energy.
%The fraction for mesons is 21\%  if only the $L=1$
%mesons are included (as in JETSET). 

Using the $T$ value  obtained from the fit, the ratio $V/(V+P)$
can be calculated according to the mass difference of vector and
pseudoscalar mesons. We obtain a value of 0.38, 0.44, 0.65 and 0.72
for  u(d)-, s-, c- and b-mesons respectively, which are in good
agreement with the measurements.
%if in the case of the D and B mesons
%the correction due to contributions from orbitally excited states
%is neglected. As will be shown in Table~\ref{bchadron} in the 
%following section, this correction
%is small in the case of the D and B mesons. 

% Fig.1: not possible to describe all data with a single exp fct
% compare with other approaches

\vspace*{-0.1cm}
\section{Predictions of heavy flavor composition}
\vspace*{-0.1cm}
Using the fit results obtained with light flavored hadrons,
the relative production rate of heavy flavored hadrons can be
predicted by our method. The absolute value of the rates can be
determined by normalizing the sum of the rates  to the 
$\Gamma_{c\bar{c}}/\Gamma_{had}$ or
$\Gamma_{b\bar{b}}/\Gamma_{had}$,
which can be calculated by the Standard Model. We use all 
heavy flavored hadrons in the JETSET~7.4 table, which includes
orbitally excited mesons with $L=1$, such as $\mbox{D}^{**}$ 
and $\mbox{B}^{**}$, but no orbitally excited heavy flavored baryons 
(for more discussions see Ref.~\cite{peippe}).

\begin{table}[t]
\vspace*{-0.4cm} 
\caption{Predictions of average production rates for heavy flavored hadrons, 
         compared with LEP data.}
\begin{center}
\vspace{-0.3cm}
{\scriptsize
\begin{tabular}{|lll|} \hline   
Hadrons           & Prediction & Measurement       \\ \hline
$\mbox{D}^{0}$    &   0.242    & $0.221\pm 0.012$    \\
$\mbox{D}^{+}$    &   0.092    & $0.087\pm 0.008$    \\
$\mbox{D}^{*+}$   &   0.114    & $0.088\pm 0.006$   \\ 
$\mbox{D}_{s}^{+}$&$0.054(\pm 0.005)$~$^{a)}$ & $0.041\pm 0.008$   \\ 
$\Lambda_{c}^{+}$ &   0.026    & $0.037\pm 0.009$   \\ 
$\mbox{B}^{0}$    &   0.091    & $0.097\pm 0.026$   \\ \hline
$\mbox{D}^{*}/(\mbox{D}^{*}+\mbox{D})$~$^{b)}$ 
                  &   0.66     & $0.56\pm 0.4$      \\ 
$\mbox{B}^{*}/(\mbox{B}^{*}+\mbox{B})$
                  &   0.70     & $0.75\pm 0.04$     \\ 
\multicolumn{3}{|l|} {$<\mbox{\scriptsize D}^{0}_{1}+
   \mbox{\scriptsize D}^{*0}_{2}>_{c}\mbox{\scriptsize Br}
   (\mbox{\scriptsize D}^{**0}\rightarrow \mbox{\scriptsize D}^{*+}\pi^{-}) 
   /<\mbox{\scriptsize D}^{*+}>_{c} $~$^{b)}$ }\\
                  &   0.124     & $0.10\pm 0.03$    \\ 
$\mbox{B}^{**}_{u,d}/\mbox{B}_{u,d}$ 
                  &   0.39     & $(0.24\pm 0.03) f$~$^{c)}$    \\ 
$(\mbox{B}_{1}+\mbox{B}_{2}^{*})_{u,d}/\mbox{B}_{u,d}$    
                         &    0.228   & $0.216\pm 0.033$    \\ 
%$(\mbox{B}_{1}+\mbox{B}_{2}^{*})_{s}/\mbox{B}_{s}$    
%                        &    0.22    & $0.175\pm 0.052     \\ 
$(\mbox{B}_{s1}+\mbox{B}_{s2}^{*})/\mbox{B}^{**}_{u,d}$ 
                         &   0.110    & $0.142\pm 0.055$    \\ 
$(\mbox{B}_{s1}+\mbox{B}_{s2}^{*})/\mbox{B}^{+}$    
                         &    0.086   & $0.052\pm 0.016$    \\ \hline
$\mbox{D}^{0}$/c-jet     &   0.593    & $0.570\pm 0.046$    \\ 
$\mbox{D}^{+}$/c-jet     &   0.237    & $0.249\pm 0.026$    \\ 
$\mbox{D}^{*+}$/c-jet    &   0.272    & $0.241\pm 0.015$    \\ 
$\mbox{D}_{s}^{+}$/c-jet &$0.101(\pm 0.025)$~$^{a)}$  & $0.128\pm 0.027$  \\ 
$\mbox{B}^{0}_{s}$/b-jet &$0.108(\pm 0.030)$~$^{d)}$     
                                      & $0.122\pm 0.031$    \\ 
$\Lambda_{c}^{+}$/c-jet  &   0.069    & $0.076\pm 0.044$    \\ 
$\Lambda_{b}$/b-jet      &   0.073    & $0.076\pm 0.019$~$^{e)}$    \\ 
$(\Sigma_{b} +\Sigma_{b}^{*})$/b-jet
                         &    0.060~$^{f)}$  
                                      & $0.048\pm 0.016$    \\ 
%$\mbox{B}^{**}_{u,d}$/b-jet 
%                         &    0.32    & $0.26\pm 0.05$     \\ 
$\mbox{D}_{s1}$/c-jet    
                         &    0.012   & $0.016\pm 0.006$    \\ 
$(\mbox{D}_{s1}+\mbox{D}_{s2}^{*})$/c-jet    
                         &    0.028   & ~~~~~---            \\ 
$(\mbox{B}_{s1}+\mbox{B}_{s2}^{*})$/b-jet    
                         &    0.035   & $0.037\pm 0.012$    \\ 
$(\mbox{D}_{1}+\mbox{D}_{2}^{*})_{u,d}$/c-jet    
                         &    0.170   & $0.173\pm 0.053$    \\ 
$(\mbox{B}_{1}+\mbox{B}_{2}^{*})_{u,d}$/b-jet    
                         &    0.188   & ~~~~~---            \\ 
$\mbox{D}^{**}$/c-jet    &    0.38    & ~~~~~---            \\ 
$\mbox{B}^{**}$/b-jet    &    0.38    & ~~~~~---            \\ 
c-baryon/c-jet           &    0.089   & ~~~~~---            \\ 
b-baryon/b-jet           &    0.091   & $0.115\pm 0.040$    \\ \hline
\multicolumn{3}{l} {$^{a)}$ for 
                        Br($\mbox{D}^{**}_{s}\rightarrow \mbox{D}_{s}X) =0.5(\pm 0.5)$} \\
\multicolumn{3}{l} {$^{b)}$ excluding D, $\mbox{D}^{*}$ and 
                        $\mbox{D}^{**}$ from B decays}\\
\multicolumn{3}{l} {$^{c)}$ $f=1$--$2$ (see Ref.~\cite{peippe})}\\
\multicolumn{3}{l} {$^{d)}$ for 
                        Br($\mbox{B}^{**}_{s}\rightarrow \mbox{B}_{s}X) =0.5(\pm 0.5)$} \\
\multicolumn{3}{l} {$^{e)}$ for 
                        Br($\Lambda_{b}\rightarrow \Lambda_{c}^{+}l^{-}\bar{\nu}X) =0.1$} \\
\multicolumn{3}{l} {$^{f)}$ $=0.052$ if  
      $m_{\Sigma_{b},\Sigma_{b}^{*}}-m_{\Lambda_{b}}$ measured in Ref.~\cite{dsb} are used}
\end{tabular}
}
\end{center}
\label{bchadron}
\vspace*{-0.7cm} 
\end{table}

As an example we consider the heavy flavor composition at  LEP energy.
Our predictions, together with the corresponding measurements~\cite{peippe,new},
are listed in Table~\ref{bchadron}. 
There is a  good agreement  between the predicted and measured values.
The main uncertainty on the prediction is due to the limited knowledge
on properties of orbitally excited heavy flavored hadrons, such as mass, decay 
modes, and branching ratios~\cite{peippe}. 
%The calculated fractions c-baryon/c-jet and b-baryon/b-jet could be 30-40\% higher 
%if orbitally excited c- and b-baryons could be included 
%in the calculation~\cite{peippe}, which are however barely known.
%
Without taking into account the production of  orbitally excited c- and b-baryons,
the fractions c-baryon/c-jet and b-baryon/b-jet are calculated to be around 9\%. 
The calculated values would be 30-40\% higher~\cite{peippe} if orbitally excited c- 
and b-baryons could be included, which are however barely known. 
%From our analysis we obtain the
%fraction of light flavored mesons and baryons with orbital excitation both to be 
%30-40\%,  and  a very similar value for heavy flavored mesons. 
%Therefore one could expect a similar value also for heavy flavored baryons, 
%which would increase the total baryon fraction by 30-40\% relative. Since baryons 
%contribute  about 10\% of the total primary hadrons, uncertainties on their rates
%have only a small effect on the predicted values for mesons. 

\section{Conclusions}
\vspace*{-0.1cm} 
We have shown that the production rates of light flavored mesons and baryons 
in  $\mbox{e}^{+}\mbox{e}^{-}$ annihilation, which span a range of four
orders of magnitude,  can be described by a simple approach
based on the idea of string fragmentation. 
Data at different center-of-mass energies can be described simultaneously
apart from a normalization factor, which reflects the rise of multiplicities 
with increasing energy. From the fit we determine 
the strangeness suppression factor to be $0.29\pm 0.02$. 
Applying to the heavy flavor production, 
we find that our predictions are in good agreement with data.
A comparison of our approach with other recently proposed approaches 
can be found in Ref.~\cite{peippe}.

The proposed  approach may provide insight into the hadron production mechanism.
To further check the approach,
precise data on the production of the decuplet baryons and 
orbitally excited hadrons are needed. 
In particular, with better understanding of the baryon production mechanism,
the description of baryon production in our approach can be improved.

\end{document}